\documentclass{aa}
\usepackage{graphicx}
\usepackage{natbib}

\begin{document}
   \title{The Mid-IR spatially resolved environment of OH26.5+0.6 at maximum luminosity
   \thanks{Based on observations
       made with the Very Large Telescope Interferometer at
       Paranal Observatory}.}

\titlerunning{}

   \author{O. Chesneau
          \inst{1}
          \and
          T. Verhoelst\inst{2}
        \and
            B. Lopez\inst{3}
          \and
          L.B.F.M. Waters\inst{2,4}
        \and
            Ch. Leinert\inst{1}
        \and
        W. Jaffe\inst{5}
        \and
            R. K\"{o}hler\inst{1}
        \and
            A. de Koter\inst{4}
            \and
            C. Dijkstra\inst{4}
          }

   \offprints{O. Chesneau}

   \institute{Max-Planck-Institut f\"{u}r Astronomie,
   K\"{o}nigstuhl 17, D-69117 Heidelberg, Germany\\
              \email{chesneau@mpia-hd.mpg.de}
         \and
            Insituut voor Sterrenkunde, K.U. Leuven, Celestijnenlaan
   200B, B-3001 Leuven, Belgium
\and Observatoire de la C\^{o}te d'Azur-CNRS-UMR 6203,\\ Boulevard
de l'Observatoire, B.P. 4229, F-06304 NICE Cedex 4         \and
Astronomical Institute ``Anton Pannekoek'', University of
Amsterdam,\\ Kruislaan 403, 1098 SJ Amsterdam, The
Netherlands 
  \and Sterrewacht
Leiden, Niels-Bohr-Weg 2, 2300 RA Leiden, The Netherlands
             }

   \date{Received; accepted }

   \abstract{We present observations of the famous OH/IR star
OH26.5+0.6 obtained using the Mid-Infrared Interferometric Instrument
MIDI at the European Southern Observatory (ESO) Very Large Telescope
Interferometer VLTI. The emission of the dusty envelope, spectrally
dispersed at a resolution of 30 from 8~$\mu$m to 13.5~$\mu$m, appears
resolved by a single dish UT telescope. In particular the angular
diameter increases strongly within the silicate absorption band.
Moreover an acquisition image taken at 8.7~$\mu$m exhibits, after
deconvolution, a strong asymmetry. The axis ratio is 0.75$\pm$0.07 with
the FWHM of the major and minor axis which are 286~mas and 214~mas
respectively. The measured PA angle, 95$^\circ \pm$6$^\circ$ is
reminiscent of the asymmetry in the OH maser emission detected at
1612MHz by Bowers \& Johnston (1990) for this star. In interferometric
mode the UT1-UT3 102m baseline was employed to detect the presence of
the star. No fringes have been found with a detection threshold
estimated to be of the order of 1\% of the total flux of the source,
i.e. 5-8~Jy. These observations were carried out during the phase of
maximum luminosity of the star, when the dust shell is more diluted and
therefore the chance to detect the central source maximized. We modeled
the dusty environment based on the work of Justannont et al. (1996). In
particular, the failure to detect fringes provides strong constraints
on the opacities in the inner regions of the dust shell or in the
close vicinity of the star.

   }

   \maketitle
%

\section{Introduction}

The short transition phase between the end of the Asymptotic Giant
Branch (AGB) phase and the formation of a White Dwarf (WD)
surrounded by a Planetary Nebula (PN) is still poorly understood.
The drastic changes observed in the circumstellar environment of
AGB and post-AGB stars  are particularly puzzling. During the late
AGB or early post-AGB evolutionary stages, the geometry of the
circumstellar material of the vast majority of stars, changes from
more or less spherical to axially symmetric, as shown by the large
number of axisymmetric proto-PNe (e.g. Sahai 2000). As a result,
most PNe exhibit axisymmetric structures, ranging from elliptical
to bipolar, often with an equatorial waist and (sometimes
multiple) jets (Corradi \& Schwarz 1995). It is thought that pure
hydrodynamical collimation provided by dense equatorial disks or
tori (Icke et al. 1989), and/or magneto-hydrodynamical collimation
(Chevalier \& Luo 1994) can explain the development of the extreme
bipolar geometries observed. Whether these equatorial structures
can arise in a single star scenario is still strongly debated
(Bujarrabal et al. 2000).

In recent years the advent of infrared spectroscopy has improved our
understanding of the AGB and post-AGB evolutionary phase. The IRAS and
the ISO infrared telescopes have detected large amounts of dust grains
around these stars. The majority of observed circumstellar environments
show either an oxygen-rich chemistry or a carbon-rich one. AGB/post-AGB
stars dominated by O-rich dust chemistry are those where the third
dredge-up never raised the C/O ratio above 1. During the AGB phase,
mass loss drives the evolution of the star. In particular, mass loss
increases dramatically (by a factor 10 at least) towards the tip of the
AGB in what is called a superwind (Iben \& Renzini 1983) which ejects
most of the remaining envelope of the star. Some O-rich AGB stars
exhibit OH maser emission and are called OH/IR stars (Wilson \&
Barrett, 1972). Their mass-loss rates are so high that the dust shell
completely obscures the central star, and the object is observable only
at infrared wavelengths and through molecular line emission at radio
wavelengths. The nature and geometry of the superwind is still to be
settled. The geometry of the maser emission is usually well constrained
due to the combination of the spatial resolution provided by
interferometric techniques and the large extension of the maser
(usually a few arcsec). The observations of the youngest (i.e. more
optically obscured) pre-planetary nebulae (PPN) where the superwind has
just ceased suggest that asymmetries are already present. An extensive
discussion on the appearance of bipolar outflows in OH/IR stars can be
found in Zijlstra et al. 2001.

OH 26.5+0.6 (RAFGL 2205, IRAS 18348-0526) is an extreme OH/IR star
showing a large dust column density and hence a very high dust
mass loss rate. It is one of the brightest OH maser emitters (Baud
1981; te Lintel Hekkert et al. 1989; Bowers \& Johnston 1990) with
a wind terminal velocity of 15 km s$^{-1}$. Bowers \& Johnston
(1990) mapped the OH maser around the star and found a shell
radius of about 2-3~arcsec. OH 26.5+0.6 exhibits a low CO $J=1-0$
and $J=2-1$ emission (Heske et al. 1990) while at 10~$\mu$m, the
silicate absorption indicates a large dust column density and
hence a very high dust mass loss rate. The 10~$\mu$m complex is
dominated by amorphous silicate absorption which has been studied
by numerous authors. The ISO spectrum of OH26.5+0.6 has been
discussed by Sylvester et al. (1999) and Molster et al. (2002)
studied particularly the signature of the crystalline silicates.

Justtanont et al. (1994, 1996, hereafter JU96) suggested that this
star has recently undergone the superwind phase and shows evidence
of two mass-loss regimes: a superwind phase in which the mass-loss
rate is 10$^{-4}M_\odot/yr$ which started recently (t $<$150 yr),
and an earlier AGB phase with a mass-loss rate of about
10$^{-6}M_\odot/yr$. The integrated mass lost during the superwind
phase has been estimated to be 0.1~$M_\odot$.

Fong, Justtanont, Meixner, Campbell (2002) reported millimetric CO
observations which did not show any significant deviation from
spherical symmetry for the envelope of OH26.5+0.6. Nevertheless, it
must be pointed out that the source is mainly unresolved at this
wavelength. In contrast, it is one of the brightest and most asymmetric
OH maser sources known among AGB stars, with a preferential axis of
symmetry oriented approximately east-west (Baud 1981; Bowers \&
Johnston 1990).

The duration of the superwind phase depends on the mass that the star
has to lose before the envelope is small enough to sustain the
mechanism of stellar (photospheric) pulsations. For a star like OH
26.5+0.6, JU96 state that the superwind {\it began very recently i.e.
less than 150 $yr$ ago}. Radio emission in molecular lines is expected
to change less rapidly than the infrared emission at the advent of the
superwind phase. It is therefore of particular interest to study the
mid-IR spatial geometry of OH/IR stars in order to determine the onset
of asymmetries in the environment of evolved stars.

Unfortunately, the dusty environment of OH/IR stars is difficult
to resolve by single dish telescopes in the IR. Even more
complicating is the time variability of the OH/IR envelopes which
modulate their size and luminosity. OH26.5+0.6 is a long period
pulsating star whose period has been refined recently by Suh and
Kim (2002) to P=1559$\pm$7~days. Taken into account the large
variations of the mid-IR flux from this star throughout its
pulsation cycle, the published data on its spatial extent have to
be systematically placed in their temporal context. Infrared
speckle interferometry has been performed by Fix \& Cobb (1988)
close to the maximum. They provide an extension for the
circumstellar dust shell at 9.7 $\mu$m (within the strong silicate
absorption) at maximum of 0.5"$\pm$0.02", while outside this
feature (at 8~$\mu$m) the shell remained unresolved by their
experiment (at most 0.2"). They have also resolved the environment
using the broad N band filter near phase 0.6 with a detected FWHM
of about 0.3" (Cobb \& Fix (1987). Some asymmetries have been
reported by Mariotti et al. (1982), Dyck et al. (1984), Cobb \&
Fix (1987) and Fix \& Cobb (1988), Starck et al. (1994). However,
the reported asymmetries are within the estimated error bars of
the measurements and all together the results are somewhat
inconclusive, and sometimes contradictory.

The Mid-Infrared Interferometric Instrument MIDI attached to the
Very Large Telescope Interferometer (VLTI) is able to provide a
spatial resolution in the mid-infrared ranging from the one
provided by single-dish 8m telescope (about 300~mas) to the one
provided by interferometric technique (about 5-10~mas). MIDI can
also disperse the light with a spectral resolution of 30 through
the entire N band which makes it a unique instrument particularly
adapted for the study of dusty environments. We used the 102~m
baseline between the telescopes Antu (UT1) and Melipal (UT3) to
observe OH26.5+0.6 for the first time.

In Section \ref{sec:datred} we describe the observations and the data
reduction procedures, divided into three parts: (i) the single dish
acquisition images (Sect.~\ref{sec:imred}), (ii) the spatial and
spectral information of the spectra (Sect.~\ref{sec:specred}), and
(iii) the interferometric signal (Sect.~\ref{sec:fringes}). In
Sect~\ref{sect:model} we model the observations using a spherically
symmetric dust model. Finally, in Section~\ref{sec:discussion} we
discuss the results of our model fitting.


\begin{table}[h]
\label{tab:journal} \vspace{0.1cm}
 \caption{Journal of observations: acquisition images}
 \begin{tabular}{llcrr}
 \hline\hline
 Star & Name & Time &Frames & t$_{exp}$ \\
 \hline
HD 168454&PSF1&06:02:09 &
2000 & 20s\\
HD 168454&PSF2&06:03:49&2000 & 20s\\
HD 168454&PSF3&06:07:34&2000 & 20s\\
HD 168454&PSF4&06:08:39&2000 & 20s\\
HD 168454&PSF5&06:14:40
&2000 & 20s\\
HD 168454&PSF6&06:15:50&2000 & 20s\\
OH26.5+0.6&star1&06:56:24 & 10000 &100s\\
OH26.5+0.6&star2&07:00:02& 5000&50s\\
OH26.5+0.6&star3&07:03:46 & 15000&150s\\
OH26.5+0.6&star4&07:07:30 &15000&150s\\
HD 177716&PSF7&08:03:11&
2000 & 20s\\
HD 177716&PSF8&08:04:23& 2000 & 20s\\

\hline
  \end{tabular}
 \end{table}

\section{Observations and data reduction}
 \label{sec:datred}

OH26.5+0.6 was observed with MIDI (Leinert, Graser et al. 2003a
and 2003b), the mid-infrared recombiner of the VLTI. The VLTI/MIDI
interferometer operates as a classical Michelson stellar
interferometer to combine the mid-IR light (N band, 7.5 -
14~$\mu$m) from two VLT Unit Telescopes (UTs). The observations
presented here were conducted in the night of the 14th of June
2003, the UT1 and the UT3 telescopes were used, separated by 102~m
with the baseline oriented 40$^\circ$ (East of North).

The observing sequence, described extensively in Przygodda et al.
(2003), is summarized hereafter. The images have been recorded
using the MIDI star acquisition modes called Default\_Chop and
Acquisition\_chop with the 8.7~$\mu$m filter. The
Acquisition\_Chop mode is the first template used after the
pointing to test if the target is within the MIDI Field Of View
(FOV) (diameter of about 3") and to perform a fine pointing. The
chopping mode (f=2Hz, angle -90 degree) is used to visualize the
star, which is not perfectly centered in the first image, and
centered in a second step. The number of frames recorded per image
is generally about 2000 and the exposure time is by default 4~ms
in order to avoid background saturation. If the result of the
template is not satisfactory, the procedure is started again. It
must be pointed out that no nodding sequences are performed, the
sky being removed by chopping only. For some stars for which the
coordinates are not well-defined, it might be difficult to get the
star directly in the MIDI FOV at the first attempt. This was the
case for OH26.5+0.6. Therefore, the Default\_Chop is used instead.
In this mode, images are recorded only for visualization and the
pointing is done by 'hand' between, or sometimes during exposures.
In this mode, the number of frames is larger, 15000 frames in our
case. The cycle rate is close to 10~ms, so we recorded 15000
frames in about 2.5~min. We stress that MIDI is not intended to be
an imager instrument but a long-baseline interferometer. Therefore
the majority of the targets are totally unresolved by a single 8m
telescope, providing a wealth of instrumental Point Spread
Function (PSF) images. The PSF files were recorded with the
Acquisition\_Chop mode and contain 2000 frames (20~s).

In the following section we present the deconvolution treatment
applied to the acquisition images.



\subsection{Images}
\label{sec:imred}

The data used to obtain a deconvolved image of OH26.5+0.6 are
summarized in Table~\ref{tab:journal}. The observations were recorded
during the acquisition process and the source location within the field
of view can be different for each file. The PSFs are generally well
centered except for PSF1. Star1 was very far from the FOV center, and
the quality of the deconvolution using this observation is very low
(but note that the results are consistent with the other measurements).

Numerous observations of two PSFs (HD~168454 and HD~177716) have
been performed before and after the star acquisition. HD~168454 is
a bright K3IIIa star exhibiting an IRAS 12~$\mu m$ flux of 62~Jy
(the IRAS flux of OH26.5+0.6 is 360~Jy). HD~177716 is a K1IIIb
which has been observed by IRAS with a flux of 26.9~Jy. There is a
Cohen template available from the ISO primary calibration
database\footnote{http://www.iso.vilspa.esa.es/users/expl\_lib/ISO/wwwcal/}
(Cohen et al. 1999). The visual seeing during the HD~168454
exposures was $\sim$0$\farcs$4, during the OH26.5+0.6 exposures it
was $\sim$0$\farcs$5 and during the HD~177716 around 0$\farcs$6.
The airmass of the three targets ranges between 1 to 1.16. The
pixel size on the sky is 98mas. This scale factor has been defined
from the MIDI observations of close visual binaries.

The deconvolution has been performed using the Lucy-Richardson
algorithm (1974) embedded in the IDL astrolib package developed by
NASA. Choosing the right iteration number for the Lucy-Richardson
algorithm is always a difficult task. Our goal is clearly not to
perform the 'best' deconvolution possible but to increase the spatial
resolution of the image which is well resolved by the UTs. The number
of iterations used was between 40 and 60. The levels where the
different deconvolved images begin to disagree among each other are
between 0.3\% to 1\% of the maximum flux of the image, depending on the
quality of the measurement. The level of the differences between PSF1
to PSF6 is about 0.3\%. The level of the differences between the PSF of
HD~168454 and the ones from HD~177716 can reach 2\% for an individual
deconvolution but is generally 1\%. PSF7 and PSF8 are quite different,
with a level of residuals reaching 1.5\%.

\begin{table}[h]
\label{tab:deconvstat} \vspace{0.1cm}

 \caption{Image size statistics}
\begin{center}
 \begin{tabular}{l|c|c|c|c}
 \hline\hline
 Name & FWHM X & rms & FWHM Y & rms  \\
 &(mas)&(mas)&(mas)&(mas)\\
\hline

PSF4&  148 &   24  &  148  &  24\\
PSF6&150 &   28  &  142  &  20\\
PSF8&160 &   26   & 166   & 22\\
 \hline
star1& 214 &   8 &     286 &   18\\
star2& 210 &   2 &     292 &  14\\
star3 & 218 &   4 &    296 &   16 \\
star4 & 212 &   2 &     268 &   12 \\
 \hline
\end{tabular}
\end{center}
 \end{table}

\begin{figure}
  \centering
\includegraphics[height=8.cm]{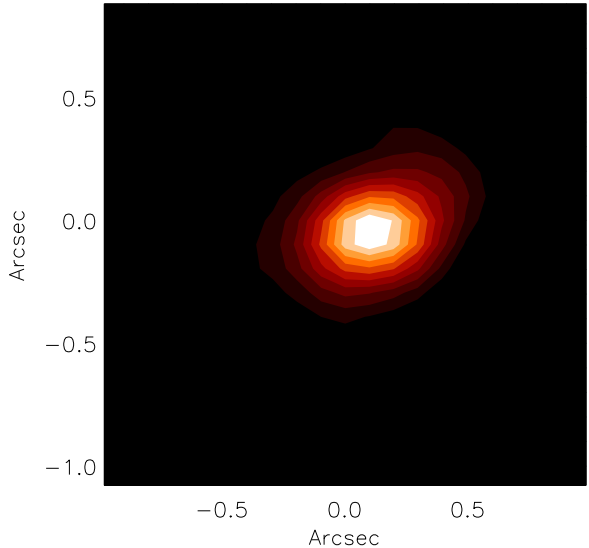}
  \caption{Contours of the mean of star 3 and star 4 deconvolved images.
The contour levels are linearly spaced for the double square root of
the image I$^{1/4}$. The last contour is equivalent to 25\% of the
maximum of I$^{1/4}$, i.e 0.4\% of the maximum of I. The three last
contours are the most susceptible to reconstruction artifacts. The
North is up and the east to the left.}
  \label{fig:meanim}
\end{figure}

The statistical properties of the PSFs have been carefully
inspected and a sub-set of good ones (5 over 8) have been selected
for deconvolution. Each data set was deconvolved using the good
PSFs and we examined the statistics of their geometrical
characteristics. A mean deconvolved image was created using the
mean image of star3 and star 4 which were located at the same
place in the detector. This reconstructed image is shown in
Fig.~\ref{fig:meanim} to illustrate the quality of the reduction
process. The position of the target for the two other files is
different and we did not attempt any shift-and-add procedure to
create a single mean image since the useful information is
extracted from a fitting procedure of the individual images. We
performed a 2D gaussian fit for each mean deconvolved image which
provides the image position, extension and the angle of the long
axis on the detector frame. Table~2. presents the statistics of
this 2D gaussian procedure and Table~3. presents the mean
parameters of the deconvolved images.

\begin{table}
\label{tab:resu} \caption{Deconvolved image parameters}
\begin{center}
\vspace{0.1cm}
\begin{tabular}{l|c|c}
\hline
Parameter & Mean & RMS\\
\hline
Mean radius & 240 mas &$\pm$14\\
Mean X axis& 214 mas &$\pm$4\\
Mean Y axis&286 mas&$\pm$6\\
Mean ratio& 0.75& $\pm$0.07 \\
Mean PA angle& 95$^\circ$ &$\pm$6$^\circ$\\
\hline
  \end{tabular}
  \end{center}
 \end{table}

\begin{figure}
  \begin{center}
\includegraphics[height=7.cm]{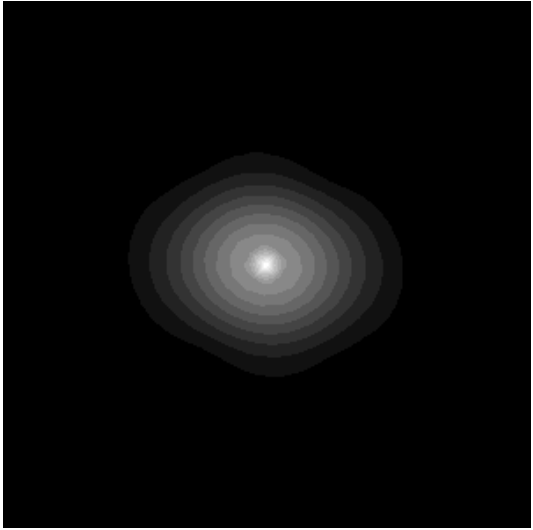}
 \end{center}
\caption[]{ Contours of the reconstructed L image from the speckle
observations at the ESO 3.6~m telescope
  on La Silla (courtesy of Starck et al.).
The contours levels are linearly spaced for the square root of the
image I$^{1/2}$.}\label{fig:speckle}

\end{figure}

We are confident that the star is indeed resolved at 8.7~$\mu$m
and this fact is settled definitely by looking at the FWHM of the
spectra (Sect.~\ref{sec:specred}). However, it is necessary to
carefully check whether the asymmetry of the image is real or not.
The image asymmetry is strong and detected with a large confidence
but the PA angle is almost coincident with the chopping direction.
Several checks were performed to ensure this detection:
\begin{enumerate}
\item First, as a comparison, the ratio between Y and X extension for
the PSFs is 0.997$\pm$0.05. Moreover the angle of the 2D Gaussian used
for the fit fluctuates randomly and there is no indication in the PSF
files that the chopping had any influence on the PSF's shape, i.e. that the
chopping flagging was uncertain during the exposures.

\item Second, the star is very bright and we have tested the
deconvolution process in some carefully chosen individual frames (4~ms
exposure) taken in the middle of the chopping cycle. The asymmetry is
already detectable with a SNR larger than 5 in the best quality frames.

\item Third, we have checked in the literature whether such an asymmetry
could have been detected in NIR by speckle interferometry in the past.
Some asymmetries have indeed been reported by Mariotti et al. (1982),
Cobb \& Fix (1987) and Fix \& Cobb (1988) in L, M and N bands. However
the axis ratios detected are not convincing, usually within the
estimated error bars of the measurement and not free of any bias as
pointed out by Fix \& Cobb (1988). Other speckle measurements in the L
and M bands are reported by Starck et al. (1994) based on observations
carried out with the 3.6m telescope of ESO/La Silla at pulsation phase
0.22 (JD=2,448,429), i.e. close to maximum luminosity. A strong
asymmetry is detected in the L band with a N-S/E-W ratio of the order
of 0.82$\pm$0.03 after removing the unresolved object (Starck, private
communication). It must be stressed that this measurement has been
performed by using more than 5 orientations on the sky preventing any
direction dependent bias. The agreement between their reconstructed L
band image and our 8.7~$\mu$m image is convincing as shown in
Fig.~\ref{fig:speckle}.

\item Fourth, surprisingly the asymmetry reported in this paper
is correlated with the strong one reported at 1612MHz by Bowers \&
Johnston (1990) at a much larger scale (few arcsecs). {\bf Even
more interesting, they reported a rotation of OH26.5+0.6's shell
at low velocity (2-3 km.s$^{-1}$) for which the projected axis is
oriented along the north-south direction.} This axis is aligned
with the minor axis of the L and 8.7~$\mu$m images and the
consequences of such a correlation will be discussed more
extensively in Sect.~\ref{sec:discussion}.
\end{enumerate}

Based on the above considerations, we are convinced that the measured
flattening is real. The data are not affected by any bias influencing
the shape of the resulting images and this asymmetry has also been seen
in other independent data sets.

\subsection{Spectrum}
  \label{sec:specred}

The photometry extracted from UT1 and UT3 is intended to calibrate
the recorded fringes. Two photometric files are recorded for each
target. In the first file, one shutter only is opened
(corresponding to UT1) and the flux is then split by the MIDI beam
splitter and falls onto two different regions of the detector. The
same procedure is then applied with UT3. The data used to get
photometrically calibrated spectra and fringes of OH26.5+0.6 is
listed in Table~4..

An independent calibration is performed for the individual spectra from
each part of the detector and for each telescope. The first step is to
read in the photometric data sets, average the frames on the target and
the frames on the sky, and subtract the average sky frame from the
average target frame. The position of the spectrum is then measured
column-wise by searching for peaks that are sufficiently high above the
background fluctuations. The result is the position and width of the
spectrum as a function of wavelength.

We use HD177716 as absolute flux calibrator (Cohen et al. 1999),
taking into account differences in air mass between calibrator and
OH26.5. Then, the calibrated spectra are combined in order to
provide a high SNR spectrum. The shape of the spectra from the
same telescope agree within 1-2\%, but the spectra from two
different telescopes can vary by about 5\%. This is due to
different optical paths which are intrinsically different during
the early use of the MIDI instrument with the VLTI (poor pupil
transfer). This defines the limit of relative error in the shape
of the spectrum (pixel to pixel and in terms of slope) which is
below 5\%. This limit has also been checked by extracting the
spectra of several spectrophotometric calibrators observed by MIDI
in several observing runs. The temporal flux variations are the
dominant source of error for the absolute flux calibration, and
variations of 5-20\% or even more are routinely observed in the N
band. During the night of the OH26.5+0.6 observations, the
photometric errors were limited to 8\%.

\begin{table}[h]
\label{tab:journalfringe} \vspace{0.1cm}
 \caption{Journal of observations: fringes (Frg) and photometric files (Phot)}
 \begin{tabular}{llccc}
 \hline\hline
 Star & Telescope & Time &Frames & File \\
 \hline
HD 168454&UT1&06:40:47&3000 & Phot\\
HD 168454&UT3&06:43:12&3000 & Phot\\
OH26.5+0.6&UT1/UT3&07:13:36 & 12000 &Frg\\
OH26.5+0.6&UT1/UT3&07:20:09 & 9000 &Frg\\
OH26.5+0.6&UT1/UT3&07:23:46 & 9000 &Frg\\
OH26.5+0.6&UT1&07:27:51& 3000&Phot\\
OH26.5+0.6&UT3&07:29:51 & 3000&Phot\\
HD 177716&UT1&08:18:44&3000 & Phot\\
HD 177716&UT3&08:20:48& 3000 & Phot\\
\hline
  \end{tabular}
 \end{table}

\begin{figure}
  \centering
        \includegraphics[height=6.5cm]{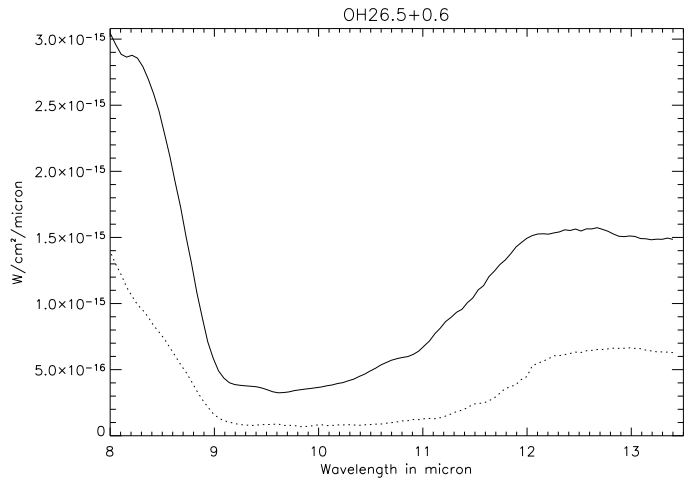}
  \caption{Calibrated spectrum from MIDI (solid line) corresponding to the mean
flux from UT1 and UT3. The flux is almost 300\% higher than the
flux observed by ISO, shown with a dotted line. The ISO data have been
recorded during a minimum of the lightcurve.}
  \label{fig:spectrum}
\end{figure}

\begin{figure}
  \centering
    \includegraphics[height=6.5cm]{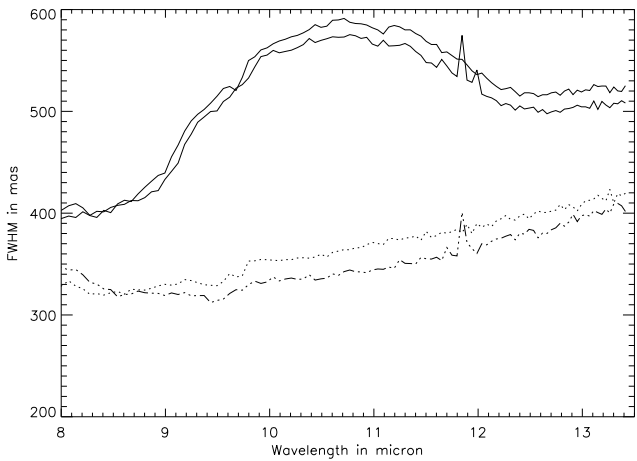}
  \caption{FWHM of the star spectrum from UT1 (solid lines) throughout the wavelength range compared
to the FWHM of the calibrator spectrum of HD 168454 (dotted
lines). There are two lines per target because the MIDI beam
splitter is inserted and the light falls onto two different
regions of the detector.}
  \label{fig:fwhm}
\end{figure}

\begin{figure}
  \centering
        \includegraphics[height=6cm]{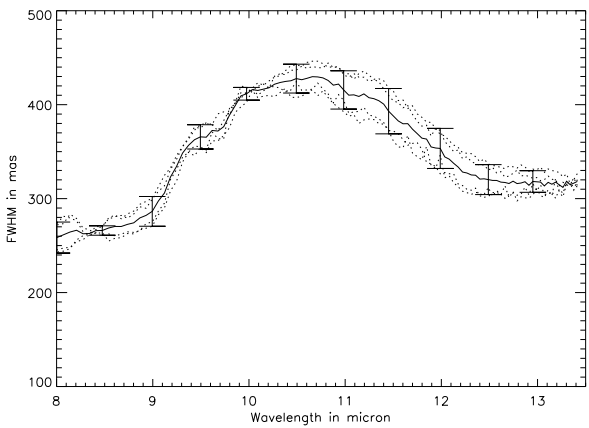}
  \caption{Mean of FWHM curves of the star deconvolved spectra from UT3
  (solid lines) and the individual
  deconvolution using different calibrators (dotted lines).
  The errors bars of the figure (spaced by 0.5 $\mu$m intervals)
  represent only the scatter of the measurements.}
  \label{fig:fwhmdeconv}
\end{figure}

We studied the spatial extension of the spectra in the direction of the
slit in order to check if the shell of OH26.5+0.6 is spatially resolved
at all wavelengths. A 1D gaussian fit was performed for each column of
each spectrum from the target and the calibrators. The PA angle of the
slit at 72$^\circ$ is close to the PA angle of the major axis detected
in the deconvolved image at 95$^\circ$. In Fig.~\ref{fig:fwhm} we see
that OH26.5+0.6 is well resolved by the 8m telescope. Moreover the star
is much larger in the silicate band. We
stress here that there was no image sharpening applied, this result
being directly extracted from the mean MIDI spectra.

In order to constrain the true size of the object in the slit
direction, we performed a deconvolution on each of the 4 spectra
available, two for each telescope. A 1D deconvolution using the
Lucy-Richardson algorithm is performed column by column using a
normalized column from the calibrators as PSF. The same number of
iterations is applied for all the wavelengths. There are systematic
differences between the shapes provided by UT1 and UT3 which can be
attributed to differences of the optical quality of the different light
paths.


\subsection{Fringes}
\label{sec:fringes}

Fringes were searched for by repeatedly scanning a large range in
optical path difference between the two telescopes, using a small
instrumental delay line (see Leinert et al. 2004 for a description of
the observing sequence). A mask is created with the average position
and width of the spectra recorded in the photometric files for UT1 and
UT3. This mask is used to extract the object data from the fringe
tracking datasets. Each frame of the fringe data, corresponding to one
individual OPD setting inside a scan, is reduced to a one-dimensional
spectrum by multiplying it with the mask and performing the weighted
integral over the direction perpendicular to the spectral dispersion.
Then the two -- oppositely phased -- interferometric output channels of
the beam combiner are subtracted from each other. This combines the
interferometric modulation of both channels into one and at the same
time helps subtracting out the background. The few dozen spectra from
each scan with the piezo-mounted mirrors are collected into a
two-dimensional array with optical wavelength and OPD as axes. The
contents of this array are column-wise fourier-transformed from OPD to
fringe frequency space.  As a rule, four of the $\approx$0.05~$\mu$m
wide wavelength (pixel) channels were added to improve the
signal-to-noise (S/N) ratio. The fringe amplitude for each optical
wavelength is then obtained from the power spectrum at the
corresponding fringe frequency.

No correlated flux has been detected with the UT1-UT3 projected
baseline of 102.4 m at a PA angle of 39.6$^\circ$. The atmospheric
conditions and the data recorded during the fringe search have
been carefully checked. The seeing degraded slowly between 6h and
8h UT from 0$\farcs$4 to 0$\farcs$6 and the standard deviation of
the flux from the target pointed by the (visible) seeing
monitor\footnote{This information has been extracted from the ESO
Ambient conditions database of Paranal observatory:
http://archive.eso.org/.} increased also, affecting the
observations of the bright calibrator HD177716, but still well
below the cloud alert threshold. During all the night, the
atmosphere turbulence was quite rapid with a mean $\tau_0$=3~ms.
These atmospheric conditions while not being excellent can
although be considered as normal conditions at Paranal
observatory. Therefore the fringe detection threshold for MIDI
during the observations OH26.5+06 was nominal.

Based on the first few months of routine observations with MIDI,
we can set limits on the amount of correlated flux the instrument
is capable of detecting under average weather conditions. For
instance, a careful data reduction of the data from NGC~1068 shows
that a correlated flux can be confidently detected down to 0.5 Jy
for faint objects (Jaffe et al. 2004). For bright objects,
visibilities of the order of 1\% have been detected from the
heavily resolved Herbig star HD~100546 (Leinert et al. 2004), or
from the clumpy environment of the supergiant $\eta$ Car (Chesneau
et al. 2004). It is difficult to reach a sensitivity less than 1\%
for bright objects because the beam combination is not perfect: a
part of the noise residuals depends on the photometric noise from
the bright source. This number has to be compared to the
photometric flux integrated over OH26.5+0.6 of about 600-800~Jy.
With the 100m baseline, MIDI is sensitive to the emission coming
from any structure smaller than 10~mas exhibiting an integrated
flux larger than 6-8~Jy in this case.


\section{Modelling the circumstellar environment}
\label{sect:model}

\subsection{The approach to modeling the object}
Most of the studies on OH/IR stars and OH26.5+0.6 in particular rely on
the interpretation of the observed variable Spectral Energy
Distribution (SED) by comparing it to a synthetic SED, computed using a
radiative transfer code. In this way, one tries to separate effects of
opacity and radial structure of the wind. Unfortunately, fits to the
SED are often not unique, especially when deviations from spherical
symmetry are taken into accout.  MIDI however provides a unique
spectrally {\bf and} spatially resolved data set which puts strong new
constaints on any model for the envelope of OH26.5+0.6. Below, we
briefly outline the strategy used to fit the MIDI observations.

Despite the convincing evidence that the envelope of OH26.5+0.6 is not
spherical, we begin our analysis assuming spherical symmetry. We use
the SED observed by ISO to determine global envelope parameters. The
ISO spectrum (and unfortunately also the IRAS data) was taken close to
the minimum luminosity of the star (Suh \& Kim 2002). Therefore the
model fit provides some constraints on the physical parameters of the
dust shell close to the minimum luminosity. The model fit parameters
are compared to the parameters published by Justannont et al. (1996),
who also mainly scaled their spectrophotometric data to the minimum
phase.

As a second step, we tried to find a good fit to the MIDI data alone
(taken at maximum luminosity), i.e. without the help of any external
spectrophotometric information as performed by Suh \& Kim (2002). This
is, we tried to fit the MIDI spectrum by performing slight
modifications to the minimum light model. Our goal was to check whether
the MIDI spectrum can be fitted based on the previous model. As soon as
a satisfactory fit is reached for the MIDI spectrum we evaluate the
spatial distribution of the flux predicted by the model and compare it
with the extension at each wavelength measured by MIDI.

Clearly it is not is the scope of the paper to find the best model
for the dust shell at maximum luminosity and such a complex study
is let to a dedicated paper. Our goal is to pinpoint the kind of
constraints provided by the inclusion of the spectrum and the
spatial extension of the object in the process of model fitting
and demonstrate how new information can emerge on the dust content
close to the star.

We use the radiative transfer code, {\sc MODUST}, commonly used for the
SED fitting of this kind of stars. The radiative transfer technique
implemented in this code has been outlined by Bouwman (2001) and
the specification of grain properties, such as size and shape
distribution, is discussed in Bouwman et al. (2000).

Throughout this analysis, we will adopt a distance of 1.37\,kpc for
OH26.5+0.6. We note that distances to AGB stars are notoriously
uncertain, and OH26.5+0.6 is no exception.

\subsection{SED fit of OH26.5+0.6 at minimum luminosity}

\begin{figure}
  \resizebox{\hsize}{!}{\includegraphics{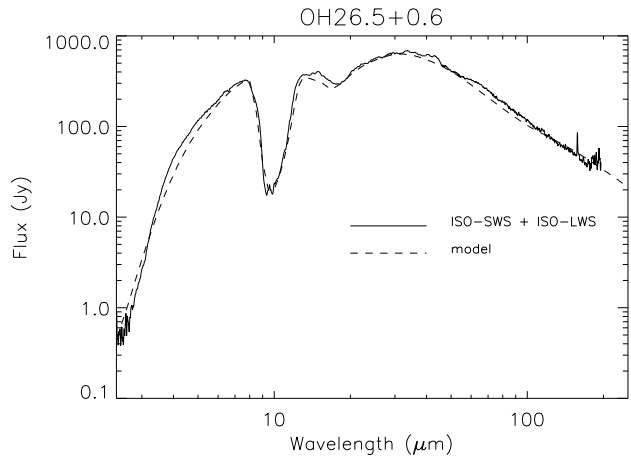}}
  \caption{Comparison between the ISO spectra (SWS + LWS) and a {\sc modust} model
  using the JU96 parameters but with a slightly reduced
  luminosity of the central star. Other differences are the use of CDE
  theory and the inclusion of metallic iron. The crystalline features
  present in the ISO-SWS observation are not
included in the model since they have little effect on the model
structure (Kemper et al. 2002)}
  \label{isosws}
\end{figure}

JU96 constructed a model for OH26.5+0.6 based on a large
collection of photometry, spectroscopy and CO-line measurements.
Their 2-component model (a very thick inner shell, due to a recent
superwind, surrounded by a tenuous AGB wind) shows that the
near-mid IR is dominated by the superwind region. Most of their
study is based on fluxes measured at minimum light, and hence
should also be applicable to the ISO-SWS spectrum,  also obtained
around minimum light (JD=2,450,368, phase 0.47)


\begin{table}[ht]
\begin{center}
\caption{Model parameters from Justtanont et al. 96 (JU96), and
the same model adapted to minimum light (the ISO data,
JD=2,450,368, phase 0.47) and maximum light (the MIDI data,
JD=2,452,804, phase 0.06), dust evaporation scenario). }
\vspace{2ex}
\begin{tabular}{l|r|r|r}
\hline
\hline
Param.    &  JU96 & Minimum & Maximum \\
\hline
T$_{eff}$ (K)   & 2200 & 2200 & 2100 \\
R$_{\star}$ (R$_{\odot}$) & 862  & 650 & 1100 \\
Dist. (kpc)   & 1.37 & 1.37 & 1.37 \\
\hline
\multicolumn{4}{c}{Superwind}\\
\hline
R$_{\rm{in}}$ (R$_{\star}$)& 7.5 & 7.5 & 20 \\
R$_{\rm{out}}$ (cm) & $8 \times 10^{15}$& $8 \times 10^{15}$& $8
\times 10^{15}$ \\
$\dot{M}$  (M$_{\odot}$/yr)  & $5.5 \times 10^{-4}$ & $5.5 \times
10^{-4}$ & $8.5 \times 10^{-4}$\\
\hline
\multicolumn{4}{c}{AGB wind}\\
\hline
R$_{\rm{in}}$ (cm) & $8 \times 10^{15}$ & $8 \times 10^{15}$ &  $8 \times 10^{15}$\\
R$_{\rm{out}}$ (cm) & $5 \times 10^{18}$& $5 \times 10^{18}$& $5
\times 10^{18}$ \\
$\dot{M}$  (M$_{\odot}$/yr)  & $1. \times 10^{-6}$ & $1.4 \times
10^{-5}$ & $1.4 \times 10^{-5}$

\label{iso_params}
\end{tabular}
\end{center}
\end{table}

The superwind hypothesis is confirmed by the ISO-SWS spectrum,
which shows very little far IR flux w.r.t. the depth of the 10~$\mu$m
feature. Assuming a density distribution going as r$^{-2}$ this can be
modelled only by cutting the shell fairly close to the star (at a few
hundred stellar radii instead of a few thousand).

Looking at the shape of the 9.7~$\mu$m feature, we can already make
some improvements on the composition of the dust. From its width and
the location of the minimum, we conclude that CDE theory (CDE,
Continuous Distribution of Ellipsoids) is to be preferred over
spherical dust particles. Furthermore, there is also strong evidence
for the presence of metallic iron, as is the case for OH127.8+0.0
(Kemper et al. 2002): the slope in the (near-)IR (4-8~$\mu$m) cannot
be explained without it. It is worth noting that the flux blocked by
the metallic iron in the near-IR emerges again in the Mid to
Far-IR. Hence, the amount of Fe will significantly influence the
optimum value of the other shell parameters.

The luminosity used by JU96 is by far too high for the epoch of the
ISO-SWS observations. However, we do obtain a satisfying fit by
reducing the stellar radius to 650 R$_{\odot}$ and keeping the outer
radius at 8$\times 10^{15}$ cm.  The comparison model vs. ISO-SWS
spectrum is shown in Fig.\,\ref{isosws}. The general shape is
approximately good. Most of the discrepancies can be attributed to the
lack of crystalline dust in our model. The crystalline features present
in the ISO-SWS observation are probably due to a few percent of
enstatite and forsterite but since these do not influence significantly
the model structure (their opacities are very similar to those of the
amorphous material, Kemper et al. 2002), we do no detailed fitting of
their spectral features.

\subsection{Attempts to account for the MIDI data}

\begin{figure}
  \resizebox{\hsize}{!}{\includegraphics{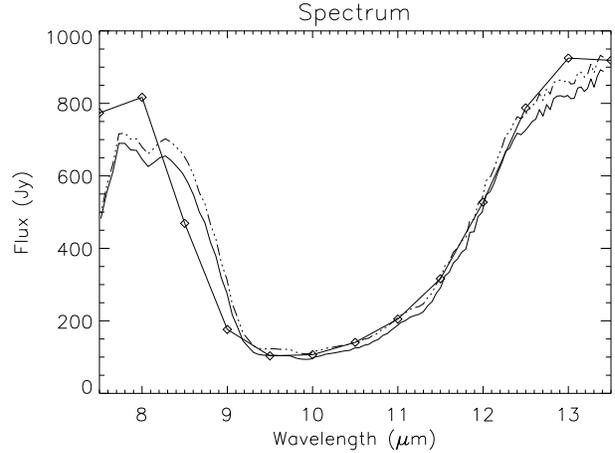}}
  \caption{Comparison between the MIDI spectrum (original: solid line,
  dereddened for IS extinction: dashed line) and the spectrum
  resulting from our ISO-tuned model but with increased central star luminosity
  (diamonds)}
  \label{bright_iso_spectrum}
\end{figure}

\begin{figure}
  \resizebox{\hsize}{!}{\includegraphics{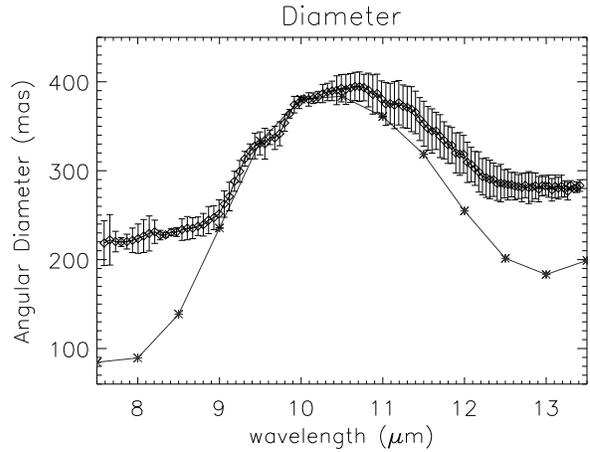}}
  \caption{A comparison between the FWHMs of the intensity profiles coming from
  the ISO-tuned model with a more luminous central star and
  the MIDI FWHMs. The predicted variation of diameter with wavelength is much
  larger than the one observed. Mainly the size at minimum optical
  thickness of the shell (at 8.5 and 13.5~$\mu$m) does not agree.
}
  \label{bright_iso_diam}
\end{figure}

Our model will have to explain the following new MIDI observations:

\begin{enumerate}
\item The MIDI N band spectrum taken close to the maximum
light of OH26.5+0.6 with the 0$\farcs$6x2$\arcsec$ slit.

\item The spatial extent of the spectrum. We limit ourselves to the
comparison of the FWHM provided by a fit of the PSF-deconvolved MIDI
spectrum by a 1D gaussian compared with a similar fit of the model
intensity profiles. The intensity distribution of OH26.5+0.6 on the sky
is likely more complex than a simple gaussian. However, it turned out
to be difficult to disentangle imperfections in the imaging quality
resulting from the many reflections in the VLTI optical train from
those intrinsic to the source, especially at lower intensity levels.
The observed FWHM is attributed to a spherical object. The slit was
oriented at PA=72$^\circ$, i.e. close to the maximum extension of the
object.

\item The negative detection of fringes by MIDI. The constraints
provided by this observational fact should not be underestimated.
During the maximum luminosity, the dust shell opacity is {\it at its
minimum}. It is quite difficult from a model of the MIDI spectrum alone
to disentangle models which are almost optically thin in the wings of
the silicate features to the ones more optically thick. The MIDI
observations definitely discard any models of OH26.5+0.6 for which the
central star is visible with a (correlated) flux larger than 3-6~Jy through the
shell at any wavelength located between 8~$\mu$m and 13.5~$\mu$m.
\end{enumerate}

The MIDI observations were done very close to maximum light
(JD=2,452,804, phase 0.06), resulting in an observed flux which is more
than twice as high compared to minimum light.
In order to fit the model at minimum luminosity
to the MIDI data at maximum luminosity, we must increase the total
luminosity to $1.7 \times 10^4$L$_\odot$.

The increase in total luminosity is simulated by an increase of the
central star diameter, still keeping the absolute value of the outer
radius of the superwind fixed. Below we confront this model with the
 MIDI FWHM observations.

\begin{figure}
  \resizebox{\hsize}{!}{\includegraphics{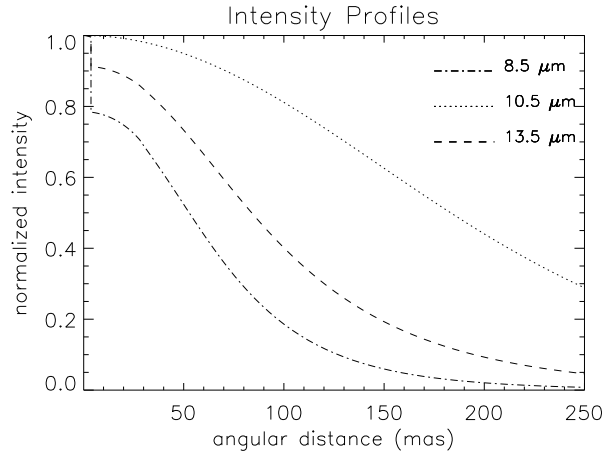}}
  \caption{Normalized intensity profiles for our model at maximum
  luminosity. In the wings of the 9.7 $\mu$m profile (at 8.5 and
  13.5~$\mu$m ), the shell optical thickness is only about 2 and thus
  the central star is visible. At 10.5 $\mu$m, the shell reaches an
  optical thickness of more than 10, resulting in the gaussian
  intensity profile.}
  \label{intprofiles}
\end{figure}

Fig.~\ref{intprofiles} shows the spatial intensity profiles according
to our ISO-tuned model adapted to the higher total luminosity at
the time of the MIDI observations. The profiles at 8.5 and 13.5~$\mu$m
correspond to a fairly low optical thickness of the shell and thus the
central star is not totally obscured. However, the amount of correlated
flux by the central star is at most a few Jansky and hence close to the
detection limit of MIDI (1~\% or 5-8~Jy). At 10.5~$\mu$m, the shell
reaches an optical thickness of more than 10, resulting in the gaussian
intensity profile.

The FWHM's determined from these intensity profiles range from 100~mas
to 370~mas (Fig.~\ref{bright_iso_diam}), clearly showing that if only
the opacity by amorphous olivines were to determine the diameter seen,
the variations with wavelength would be much larger than what is observed. The
maximum size appears to compare reasonably well with the model (though
with a slightly different wavelength of maximum) and hence is
compatible with the superwind size of JU96. More precisely, we can put
a lower limit to the size of the superwind region of 400\,mas, which
corresponds to $4 \times 10^{15}$\,cm at 1.37\.kpc.

\begin{figure}
  \resizebox{\hsize}{!}{\includegraphics{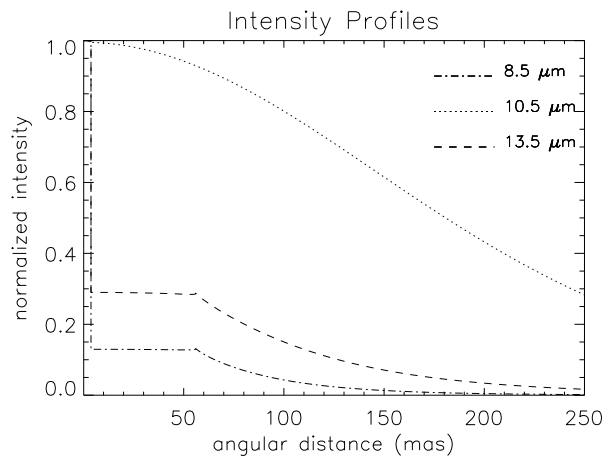}}
  \caption{Normalized intensity profiles for our model at maximum
  luminosity with an increased inner radius. For such intensity profiles, the inner radius is
  determining the observed size of the object at wavelengths of low opacity.}
  \label{dust_evap_profiles}
\end{figure}

\begin{figure}
  \resizebox{\hsize}{!}{\includegraphics{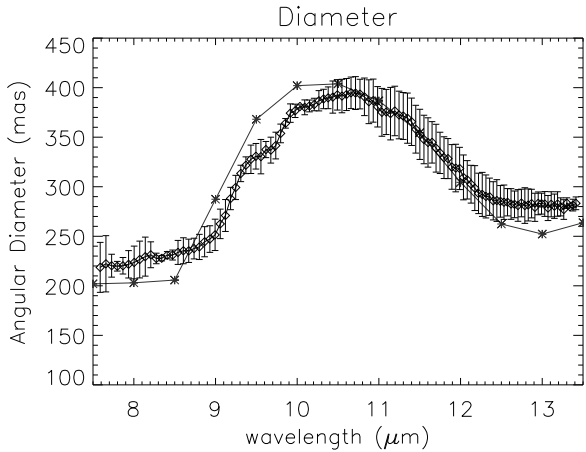}}
  \caption{The large observed radius at the red and blue sides of the
  profile can be simulated by moving the inner radius of the model far
  out, to about 20 R$_{\star}$. }
  \label{farout}
\end{figure}

The discrepancy between the spatial extent of the model and the
observed size of the dust shell near 8 and 13 $\mu$m can be resolved
if we move the inner radius of the dust shell to larger distance from
the star. The minimum size seen will be mainly determined by the inner
radius of the dust shell: at the wavelengths of low optical depth, the
intensity profile is not at all gaussian
(Fig.~\ref{dust_evap_profiles}). In this way, the observed diameter in
the wings of the 9.7 $\mu$m feature can be simulated by moving the
inner radius out to about 20 R$_\odot$ (Fig.~\ref{farout}).  For this
model, a slightly higher mass loss ($8.5\times 10^{-4}$\,M$_{\odot}$)
is needed to keep the quality of the spectral fit.

At first sight, an alternative solution would be to introduce a source
of opacity with only a modest wavelength dependence, which would
dominate over the silicate dust opacity near 8 and 13 $\mu$m. This
could either be gas-phase molecular opacity or dust. However, by doing
so the spectral fits become unacceptably poor, because the depth of
the silicate feature can no longer be reproduced.


In summary, we conclude that the spectral and spatial data of
OH26.5+0.6 can be understood in the framework of a spherically
symmetric shell, with dust components that are also shown to be
present in other OH/IR stars. The outer radius of the dust shell
agrees well with the one estimated by Justannont et al. (1996) as the
radius of the onset of the superwind. The spatial data near 8 and 13
$\mu$m force us, in the context of spherical symmetry, to move the
inner radius of the dust shell to a distance of about 20 stellar
radii. However, this results in a predicted correlated flux using the
102 meter UT1-UT3 baseline which is 5 times the upper limit imposed by
the non-detection of fringes in the interferometric signal (under the
assumption that the central star has a typical AGB temperature of
3000-4000\,K).



\section{Discussion}
\label{sec:discussion}

\subsection{Nature of the large inner dust radius}

While a large inner radius of the dust shell seems a simple solution to
our fitting problems, it is clearly not in agreement with the limits
set by the interferometric measurement. In addition, such a large inner
radius is not compatible with our current understanding of oxygen-rich
AGB dust shells: the dust temperature at the inner edge of our dust envelope is only
500-600\,K, well below what is believed to be the condensation
temperature of olivines (1000\,K). Furthermore, one can wonder whether
the region between photosphere and olivine dust shell contains other
material (refractory residuals of dust for instance, like corundum).
Given the constraint that whatever fills this region must be quite
transparent from 8 to 13~$\mu$m\footnote{This means optically thin from
7.5 to 9 micron and from 11.5 to 13.5 micron. An optically thick shell
($\tau \simeq$2-3) would ruin the diameter profile since a smaller
radius would be necessary to account for the SED.}, several hypotheses
can be formulated:

\begin{itemize}

\item{A large cavity within this inner region, could indicate that the mass loss has
decreased strongly about 30 years ago. This is compatible with the
timescales derived from the rings observed around several Post-AGB
stars, hinting at episodic mass-loss with periods of a few hundred
years (e.g.IRAS 17150-3224: Kwok, 1998;IRC+10216: Mauron \& Huggins
1999; Egg Nebula:Sahai, 1998, Marengo, 2001). However, stellar
pulsations are quite regular and have been detected over the past 30
years (e.g. Suh and Kim,2002). So, if mass loss stopped, apparently the
pulsations did not. Moreover, this model offers little opacity at 8
micron ($\tau_{8 \mu m}$ is of the order of 1), and hence the amount of
correlated flux from the central object would be of the order of 5
times the detection limit of MIDI. This puts into question the relation
between pulsations and mass loss for AGB stars in the superwind phase,}

\item{The mass loss is continuing even today and no condensation of the dust
is possible inwards of 20 stellar radii when the star is at maximum
luminosity. In fact, early dust condensation models by Sedlmayr (1989)
have predicted that the dust condenses only when the gas is extremely
super-saturated, which happens well below the glass temperature, and
would be around 800-600K. Several other studies have previously hinted
at the possibility that dust formation in AGBs does not happen close to
the star. Danchi et al. (1994) also find some examples of stars with
rather detached dust shells, corresponding to timescales of decades, so
similar to what we find. However, other stars have inner dust radii
much closer to the star.}
\item{The dust gets periodically destroyed (dust evaporation)
because the stellar luminosity changes during a pulsation cycle. Some
calculations were done by Suh and Kim (2002), but do not predict the
required amount of dust destruction to agree with our geometrical
model. It is not certain that the net amount of dust created through an
entire cycle is stable. This means that the episodical mass loss is a
matter of balancing between destruction and formation. Of course, if
the dust formation is not large enough to compensate for the
evaporation, the inner gap would increase in size over multiple cycles.}

\end{itemize}

It is not straightforward to choose between these possible
explanations, and more importantly, not one of them explains the
non-detection of fringes.

Although the SED is compatible with our
spherically symmetric model, this assumption might be strongly
violated, as suggested by the acquisition image.  It is intuitively
clear that a disk+bipolar outflow structure could account for the fitting
problems we experienced. If the MIDI slit was oriented perpendicular
to a nearly edge-on disk, the very high density (and thus opacity)
would explain both the large size in the wings of the silicate feature
and the non-detection of fringes. Because of the much larger
complexity of 2D CSE modelling, we leave an in-depth analysis of this
hypothesis to a future dedicated paper.

\subsection{Star asymmetry}
The drastic changes observed in AGB and post-AGB stars
circumstellar structure are particularly puzzling. Information on
the geometry of the AGB mass loss was first obtained from
interferometry maps of OH 1612~Hz maser emission (Booth et al.
1981; Herman et al. 1985, Bowers \& Johnston 1990). These, as well
as CO radio line maps of a large number of AGB-stars (Neri et al.
1998, Olofsson et al. 1999 and references therein), were
consistent with an over-all spherically symmetric mass loss.
Evidences concerning deviation of the geometry of OH/IR stars can
be found in the literature but the results are generally somehow
contradictory and perplexing due to the large range of masses and
evolutionary stages encompassed by this term, from embedded AGBs
to PPNs (Van Winkel, 2003). In particular, the sub-group of OH/IR
star for which no pulsation can be detected is associated with
PPNs (Zijlstra et al. 1991).

What makes the MIDI observations particularly interesting is that
JU96 demonstrated that the star entered the superwind phase
$\sim$200yr ago. This provides an extremely short upper limit to
the development of a large scale asymmetry. For a large mass star
like 0H26.5+0.6, the duration of the OH/IR phase is expected to be
of the order of 10$^4$yr, to be compared to 10$^3$yr for a
low-mass star (JU96). Increasing evidence show that AGB wind
becomes axi-symmetric at the very last stages of the AGB
evolution, and that the interaction with a fast wind from the
post-AGB object further enhances the axi-symmetry. The MIDI
observations suggest that in the case of OH26.5+0.6 the appearance
of asymmetries can occur is a fairly short time scale (Sahai et
al. 2003).

JU96 and Fong, Justtanont, Meixner, Campbell (2002) reported CO
observations which did not show any significant deviation from
spherical symmetry for OH26.5+0.6, but at most of the emission is
spatially unresolved (coming mostly from the superwind). In
contrary, the OH 1612~Hz maser emission from Baud (1981) and
Bowers \& Johnston (1990) present a clear picture of the clumpy
and asymmetric environment OH26.5+0.6. The radio shell of
OH26.5+0.6 is certainly one the most extended and least symmetric
known for OH/IR stars. The large scale environment of OH26.5+0.6
is crowded which not ease the extraction of the radio emission
sensitive to any anisotropic UV radiation field and Bowers \&
Johnston (1990) proposed that it is a likely cause for the
detected asymmetry.

{\it The crucial point is that the axis of symmetry of the present
Mid-IR objet is perfectly aligned to the large scale anisotropy
detected in radio}, which excludes a-priori a strong external
influence on the shaping of the OH maser. This correlation opens
new possibilities of interpretation. Hence, Bowers \& Johnston
(1990) detected some hints of rotation at low projected velocity
(v$_r <$3 km~s$-1$) with a rotational axis aligned with the minor
axis of the asymmetric shell.

What could be the origin of the development of such an asymmetry?
It is not in the scope of this paper to review all the mechanisms
invoked for explaining such a phenomenon and the reader is
invited to consult the review from Balick \& Frank (2002). We just
point out that OH26.5+0.6 is not a known binary, but owing to the
difficulties to study the central star of OH/IR star, this lack of
detection is not meaningful. The discussion above lead us to think
that OH26.5+0.6 particular characteristics are perhaps better
understood under the binarity hypothesis.

\subsection{Improving the model}
By using an up-to-date spherical model of a dust shell, we have
been able to fit satisfactorily the SED of the star but this model
failed to provide a direct explanation of the non-detection of any
fringes within the N band.

These difficulties point to a problem of opacities located in
regions fairly close to the star though sufficiently extended to
prevent the detection of correlated flux by MIDI. One of the
remedy is to make the radius of the central source larger so that
almost no correlated fluxes can be detected by MIDI with a
baseline as large as 100m.

An attractive solution to this problem would be the inclusion of
molecular opacities. Growing evidence of their deep effects on
interferometric measurements in the near and mid-infrared are
reported (Matsuura et al. 2002, Mennesson et al. 2002, Perrin et
al. 2004a, b, Schuller et al. 2004, Cotton et al. 2004, Ohnaka et
al. 2004). The first effect is to increase the diameter of the
central star and thus, to decrease the correlated flux. At maximum
luminosity, the expected angular diameter OH26.5+0.6 is about
8~mas (for R$_*$=1100R$_\odot$) and the correlated flux from the
central object should represent about 80\% of the stellar flux if
the star is a uniform disk. The inclusion of an optically thick
molecular envelope of H$_2$O and SiO of about 2.5R$_*$ divides
this correlated flux by 10, preventing probably its detection by
MIDI. Moreover, the star can probably no longer be modelled by a
uniform disk but by a spatially smoother flux distribution which
decreases again the correlated flux. The second effect is to
redistribute the flux from the central star to other regions of
the spectrum. Of course, if the warm molecular layers are
optically thick they will emit like a blackbody at a temperature
slightly lower than the star. The effects of this envelope on the
dust formation/destruction processes have to be carefully
evaluated and need consistent radiative transfer calculations
which are not in the scope of this paper.

The opacity of the dust shell remains an issue which could be
solved by changing the dust density without affecting its
composition or by adding species particularly absorptive between 8
and 13.5$\mu$m. This implies either a very large olivine dust
column density, either other highly absorptive dust species like
corundum. An inner shell rich in metallic iron and corundum
(Al$_2$O$_3$) could help to fit the observed extensions,
preventing the flux from the star to be detected. The corundum is
used for instance to model the opacities of the thin dust shell
around some Miras, with mass ratio of corundum to silicate ranging
from 0.6 to 0.9 and grain sizes of 0.1~-~0.2~$\mu$m (Martin-Lorenz
\& Pompeia, 2000, Ohnaka et al. 2004). However, the amount of
material needed is limited by the natural abundance of Al in the
photosphere. In our first tests, this amount needed is at the
moment unrealistically high to prevent the correlated flux from
the star to be detected, under the hypothesis that the star is a
naked photosphere of about 1100R$_\odot$ at 2100K (i.e. a uniform
disk).

Finally, for the sake of simplicity, we have put all the
discussions in the frame of a spherical object. All the codes used
to model the SED of OH26.5+0.6 have been using spherical geometry
to understand an object which is proven to be strongly flattened
in this article. {\it A very promising hypothesis is that we are
indeed observing a high olivine column density in the direction
perpendicular on the slit, i.e. that we are looking an equatorial
overdensity (or even a disk) close to an edge-on configuration,
explaining the important aspect ratio of the 8.7$\mu$m MIDI
image.}

\subsection{Envelope clumping}
The fact that no fringes have been detected from OH26.5+0.6
implies also that the dusty environment of OH26.5+0.6 is
relatively homogeneous and smooth. Most of the flux originates
from the dust shell and the absence of fringes is in great
contrast with the almost ubiquitous fringes found around the
massive star Eta Car in an area as large as
0$\farcs$6$\times$0$\farcs$6 with photometric fluxes comparable to
the ones reported in this paper (Chesneau et al. 2004). The use of
continuous dust distribution for the modelling of this kind of
environment is thus fully justified.

The pulsations are supposed to generate a strongly clumped medium
due to the shocks, but Suh et al. (1990) has shown that this
region is limited to the 3~R$_*$. The rapid outward acceleration
extending to 10-20R$_*$ should considerable smooth the dusty wind.
In the Suh et al. model, the dust condensation radius is about
6~R$_*$ depending on the pulsation phase. Even considering the
pulsation, their model of the dusty envelope is very close to a
smooth $r^{-2}$ density law. The clumpy regions embedded in the
optically thick part of the shell at 10~$\mu$m should not be
visible but some signal could be expected at 8 or 12~$\mu$m at
maximum luminosity if the dust shell is sufficiently optically
thin. We note that the clumps are probably embedded in the
putative optically thick molecular layer so that their correlated
emission would largely hidden and therefore undetectable by MIDI.

\section{Conclusion}
It has been shown that the dust model used in this article to
interpret the SED from OH26.5+0.5 has strong difficulties to
predict the large extension of the dust shell outside the silicate
absorption region with simultaneously having opacities sufficient
to render the flux from the central object undetectable by MIDI.
The conjunction of these two complementary constraints make it
necessary to pursue a deeper effort to understand the physical
processes operating at the inner regions of the dust shell.

OH26.5+0.6 is indeed a very complex object exhibiting a wide range
of physical phenomenon:
\begin{itemize}
\item an asymmetric appearance, whose axis of symmetry is probably
coincident with the axis of rotation of the star,
\item a thick dust envelope whose characteristics are modulated by
the pulsation cycle,
\item a complex inner shell where dust forms and is destructed
throughout the cycle. The contribution of corundum and metallic
iron opacities in this region are probably important,
\item a putative thick molecular envelope as encountered in many, if not all, Mira stars
which increases the angular diameter of the central star, and
decreases the apparent temperature,
\end{itemize}

The results presented in this paper are very constraining and have
to be confronted with a model able to handle consistently the
complex interplay between the pulsating central star, its
molecular atmosphere and the mechanisms of dust
formation/destruction and transport. However, such a theoretical
approach is at the moment inefficient until any confirmation of
the presence of a disk around this object allows a restricted
range of geometrical parameters.

In the course of the $\sim$1560~days of the cycle, MIDI/VLTI
interferometer will allow a continuing monitoring of OH26.5+0.6.
The observations will provide a unique view of the evolution of
the {\bf size and the shape} of the dusty envelope throughout the
entire cycle. However the phase of maximum luminosity remains the
unique opportunity to reach the most internal regions close to the
star.

High resolution observations in optical and near-infrared by means
of Adaptive Optics should also help to estimate the amount of
scattered light close to the object in order to test efficiently
the disk hypothesis.

\begin{acknowledgements} We acknowledge fruitful discussion with
Carsten Dominik and Ciska Kemper. O.C. acknowledges the Max-Planck
Institut f\"{u}r Astronomie in Heidelberg, Germany and in
particular Christoph Leinert and Uwe Graser for having given him
the opportunity to work for a motivating project within a rich
scientific environment.

\end{acknowledgements}

\end{document}